\documentstyle[preprint,prd,aps]{revtex}
\begin{document}
\draft
\preprint{
\vbox{
\halign{&##\hfil\cr
	& NUHEP-TH-96-1 \cr
	& hep-ph/9602411 \cr
	& February, 1996 \cr}}
}
\title{Comparative Study of the Hadronic Production of $B_c$ Mesons}

\author{ Chao-Hsi Chang}
\address{ China Center of Advanced Science and Technology 
(World Laboratory), Beijing, China}
\address{ Institute of Theoretical Physics, 
 Academia Sinica, Beijing 100080}

\author{ Yu-Qi Chen and Robert J. Oakes }
\address{ Department of Physics and Astronomy, Northwestern University,
	Evanston, IL 60208 }

\maketitle

\begin{abstract}
A comparative study of the full $\alpha_s^4$ perturbative QCD calculation 
of the hadronic production of the $B_c$ and $B_c^*$ mesons and the 
fragmentation approximation is presented. We examined the various subprocesses 
in detail and the distribution $C(z)$, where $z$ is twice the fraction of 
the $B_c$ or $B_c^*$ meson energy in the center of mass of the subprocess,
which proved insightful and revealed the importance of certain 
non-fragmentation contributions. We concluded that the condition for 
the applicability of the fragmentation approximation in hadronic collisions 
is that the transverse momentum $P_T$ of the  produced $B_c$ and $B_c^*$ be 
much larger than the mass of the $B_c$ meson; i.e., $P_T\gg M_{B_c}$. 
Numerical results for the cross sections at the Fermilab Tevatron are also 
presented using updated parton distribution functions with various kinematic 
cuts. 
\end{abstract} 

\vfill \eject

\narrowtext

The heavy flavored $\bar{b}c$ meson states have attracted considerable 
interest due to their interesting properties. The experimental search for 
these mesons is now under way at high energy colliders such as the LEP 
$e^+e^-$ collider and the Tevatron $\bar{p}p$ collider. Results are 
expected in the near future. An important theoretical task, relevant to 
these experiments, is to calculate the production cross sections. Since 
they are weakly bound states hadronization is relatively simple and their 
production cross sections can be accurately predicted in the framework of 
Perturbative QCD (PQCD). At LEP the production of the pseudoscalar
ground state $B_c$ and the vector meson state $B_c^*$ is dominated 
by $Z^0$ decay into a $b\bar{b}$ pair, followed by the fragmentation 
of a $\bar{b}$ quark into the $B_c$ or $B_c^*$ meson\cite{9,60}. Hadronic 
production, as first pointed out by Chang and Chen\cite{0}, is dominated
at high energies by the subprocess $gg\to B_c(B_c^*)b\bar{c}$. The 
interesting question we address here is how well the fragmentation 
approximation to the full order $\alpha_s^4$ calculation works for the 
hadronic production of the $B_c$ and $B_c^*$ mesons and in which kinematic
region.

The hadronic production of the $B_c$ and $B_c^*$ can be calculated
fully to order $\alpha_s^4$ in PQCD, which involves a difficult numerical 
calculation. Since Chang and Chen\cite{0} first presented the numerical 
results for the hadronic production, calculations have also been done 
by several other authors\cite{1,2,3,6,5,50,7}, not all of which are 
in agreement. Slabospitsky\cite{1} claimed an order of magnitude larger 
result than Chang and Chen\cite{0}. Berezhnoy, {\it   et al.}\cite{2} 
obtained a result larger than Chang and Chen\cite{0} and smaller than 
Slabospitsky\cite{1}. Masetti and Sartogo\cite{3} found a result similar 
to Berezhony, {\it   et al.}\cite{2}. More recently, Berezhony, {\it et al.} 
\cite{5} found that a color factor $\displaystyle{1\over 3}$ was overlooked 
in their previous work\cite{2}. After including this factor, their revised 
result is in agreement with Chang and Chen\cite{0}. Baranov\cite{50} 
independently also obtained results similar to Chang and Chen\cite{0}. 
Ko{\l}odziej, {\it  et al.}\cite{7} presented results using different 
parton distribution functions and energy scale from that used by Chang 
and Chen\cite{0} so it difficult to directly compare their final result. 
However, their results for the cross sections for the subprocess are similar 
to others\cite{6,5}. Therefore, we are confident that the results of the 
full order $\alpha_s^4$ PQCD calculations in Refs.~\cite{0,6,5,50,7} are 
in agreement.

An alternative way to calculate the hadronic production is to use the 
fragmentation approximation. From the general factorization theorem
it is clear that, for large transverse momentum $P_T$ of the $B_c$ or $B_c^*$, 
hadronic production is dominated by fragmentation. The calculation 
can then be considerably simplified using this approximation, as was 
first done by Cheung\cite{10}. Subsequently, the comparison between 
the full $\alpha_s^4$ calculation and the fragmentation approximation
has been discussed by several authors\cite{6,5,7}. However, very different 
conclusions have been drawn, although their numerical results are 
similar. Chang, {\it et al.}\cite{6} found that when $\sqrt{\hat{s}}$ and 
$p_T$ are small the difference between the fragmentation
approximation and  the $\alpha_s^4$ calculation is large. 
Berezhony, {\it  et al.}\cite{5} claimed that the fragmentation approximation 
breaks down even for very large $P_T$ by examining the ratio of $B_c^*$ 
to $B_c$ production. Ko{\l}odziej, {\it  et al.}\cite{7} claimed that 
the fragmentation approximation works well if $P_T$ exceeds about $5-10$ GeV, 
which is comparable to the $B_c$ mass, by investigating the $P_T$ distribution
of the $B_c$ meson.  

It is nontrivial to clarify these points since the 
full calculation is quite complicated, the dominant subprocess involving 36
Feynman diagrams; but the importance of investigating this issue is twofold:
From the theoretical perspective, it provides an ideal example 
to quantitatively examine how well the fragmentation approximation 
works for calculating the hadronic production of heavy flavored 
mesons. In this process both the full $\alpha_s^4$ contributions and the 
fragmentation approximation can each be calculated reliably. 
Experimentally, it is important to have a better understanding of the 
production of the $B_c$ mesons in the small $P_T$ region where the $B_c$ 
production cross section is the largest. The $P_T$ distribution 
decreases very rapidly as $P_T$ increases.

We carried out a detailed comparative study of the fragmentation 
approximation and the full $\alpha_s^4$ QCD perturbative calculation. 
We first studied the subprocesses carefully. By analyzing the 
singularities appearing in the amplitudes and the $P_T$ distributions 
of the subprocesses we gained very important insight into the processes. 
We then calculated the entire hadronic production cross section for 
$p\bar{p}$ (or $pp$)  collisions at the Tevatron energy. To facilitate 
a quantitative comparison we found it very instructive to examine the 
distribution in $z$, twice the energy fraction carried by the $B_c$
meson in the center of mass of the subprocess, which is experimentally 
observable. By investigating the $z$ distribution we found that the 
fragmentation approximation dominates when and only when $P_T\gg M_{B_c}$. 
We also found that it is insufficient to evaluate the validity of the 
fragmentation approximation by only examining the $P_T$ distribution. 
In fact, we found that it is simply fortuitous that the $P_T$ distribution 
in the fragmentation approximation for the case of the $B_c$ meson is 
close to the results of the full order $\alpha_s^4$ calculation when
$P_T$ does not satisfy the above condition; i.e., $P_T\gg M_{B_c}$. It 
is not sufficient for $P_T$ to simply be the same order as $M_{B_c}$, 
as we discuss below.

In order to obtain some insight into the process we first focus the 
discussion on the subprocess. At the lowest order, $\alpha_s^4$,
there are 36 Feynman diagrams responsible for the dominant gluon fusion
subprocess $g(k_1)+g(k_2)\to B_c(p) +b(q_2) +\bar{c}(q_1)$, where $k_1$, 
$k_2$, $p$, $q_1$, and $q_2$ are the respective momenta.
When the energy in the center of mass system, $\sqrt{\hat{s}}$, is much 
larger than the heavy quark mass the main contributions to the cross
section come from the kinematic region where certain of the amplitudes
in the matrix element are nearly singular; i.e., some of the quark lines 
or gluon lines are nearly on-shell and the related propagators in the 
Feynman diagrams are nearly singular. This results in the cross section for 
the subprocess coming from the lowest twist contributions being proportional 
to $\displaystyle {1\over \hat{s} }{ f_{B_c}^2\over M_{B_c}^2}$, where 
$f_{B_c}$ is the $B_c$ decay constant, with some logarithmic correction 
terms such as $\ln(\hat{s}/M_{B_c}^2)$. Now the possible singularities in 
the square of the matrix element for this subprocess must arise from the 
inverse power(s) of the following factors, or their products, which can 
appear in all the possible denominators of the quark and gluon propagators 
in the Feynman gauge:
\begin{equation}
\begin{array}{ccccc}
q_i\cdot k_j, &~~~ p\cdot k_j, 
&~~~ (\alpha_ip + q_i)^2,
&~~~ {\rm and} 
&~~~ (k_j- \alpha_1 p - q_1)^2,
\end{array}
\end{equation}
where $i,j=1,2$ and  
$\displaystyle \alpha_{1,2}= {m_{c,b}\over (m_c+m_b)}$ is the ratio
of quark masses. It is easy to see that when the $P_T$ of the $B_c$ meson 
is large only $ (\alpha_ip + q_i)^2 $ can still be small ($\sim m_i^2$). The 
fragmentation functions can then be extracted from the most singular part 
containing the inverse powers of this factor in the square of the matrix 
element. It then follows that in the large $P_T$ region the subprocess is 
dominated by the fragmentation approximation. {\em However, when the  $P_T$ of 
the $B_c$ meson is small the produced $B_c$, as well as the $b$ and the 
$\bar{c}$ quarks, can be soft or collinear with the beam.} In this region 
the square of the matrix element is highly singular because  {\em two or more} 
of the internal quarks or gluons in certain Feynman diagrams can 
{\em simultaneously} be nearly on-mass-shell. Although this region is a 
smaller part of the phase space, these nearly singular Feynman diagrams, in 
fact, make a large contribution to the cross section and dominate the small 
$P_T$ region.

Generally, in the square of the matrix element we can isolate all 
the terms which contribute to the lowest twist cross sections
using singularity power counting rules\cite{51}. When 
$\sqrt{\hat{s}}\gg M_{B_c}$ the lowest twist contributions dominate,
while the higher twist contributions are suppressed by a factor 
$\displaystyle{m^2/\hat{s}}$ and are negligible. We can decompose the 
terms which contribute to the lowest twist cross sections into two components: 
One component is the fragmentation contribution, which dominates the 
large $P_T$ region. The other is the non-fragmentation component, 
which comes from the other singular parts of the matrix element, 
those in which {\em two or more} quarks or gluons are nearly on-shell,
as discussed above. This non-fragmentation component dominates in 
the smaller $P_T$ region. The contributions of these two components are 
quite clearly distinguishable in the $P_T$ distribution of the subprocess, 
particularly at large $\sqrt{\hat s}$.  In Fig. 1 we show the $P_T$ 
distribution of the subprocess when $\sqrt{\hat{s}}=200$ GeV for both 
the full $\alpha_s^4$ calculation and the fragmentation approximation
with $\alpha_s=0.2$, $m_b=4.9$ GeV, $m_c=1.5$ GeV, and $f_{B_c}=480$ MeV.
In the fragmentation calculation, to reduce the error caused by the phase 
space integrations, we directly used the squared matrix elements\cite{9,80}, 
from which the fragmentation functions are derived\cite{9,80}, rather than 
the fragmentation functions themselves. It is easy to see in Fig. 1 that 
when $P_T$ is larger than about 30 GeV for the $B_c$ and about 40 GeV for 
the $B_c^*$, the fragmentation approximation is close to the full 
$\alpha_s^4$ calculation. However, when the value of the $P_T$ is smaller 
than about 30 GeV for the $B_c$ and about 40 GeV for the $B_c^*$, the 
deviation between the fragmentation approximation and the full calculation 
becomes large and the non-fragmentation component clearly dominates their 
production. This critical value of $P_T$ is certainly much larger than the 
heavy quark masses, or the $B_c$ meson mass; we also found that it slowly 
increases with increasing $\hat{s}$, which may indicate that there is an 
additional enhancement due to logarithmic terms such as $\ln{\hat{s}/m^2}$ 
in the non-fragmentation component compared to the fragmentation component. 
When ${\sqrt{\hat s}}$ is not very large this two component decomposition is 
less distinct, since the higher twist terms can not be ignored\cite{6,5,7}. 
In this case, the fragmentation approximation is not a very good 
approximation, there being quite a large discrepancy with the full 
$\alpha_s^4$ calculation\cite{6,5,7}. 

A similar process is the production of $B_c$ and $B_c^*$ mesons in 
photon-photon collisions, which is also instructive to examine more 
carefully.  A comparative study of the full $\alpha^2\alpha^2_s$ calculation 
with the fragmentation approximation in this process was presented 
in Ref.\cite{12,4}, where it was claimed that the fragmentation 
approximation is not valid.  There are 20 Feynman diagrams which can 
be divided into four gauge invariant subsets corresponding to various 
attachments of photons onto the quark lines; i.e., subsets I, II, III, 
and IV corresponding respectively to the attachment of both photons 
onto the $b$ quark line, onto the $\bar{c}$ quark line, one photon 
onto the $b$ quark line and the other onto the $\bar{c}$, and the 
interchange of $b$ and $\bar{c}$. Subset I is dominated by the $\bar{b}$ 
quark fragmentation into the $B_c$ when $P_T$ is large, as discussed above. 
Subsets III and IV, called recombination diagrams, can only contribute to 
the non-fragmentation component and decrease rapidly when $P_T\gg M_{B_c}$, 
as also discussed above. However, subset II is somewhat unusual. Although 
this contribution is relatively suppressed by the smaller probability for 
subsequent $b\bar{b}$ quark creation, it nevertheless gives quite a large 
contribution to the total cross section because of the enhancement of the
$c$ quark electric charge. For example, we found that when 
$\sqrt{\hat s}=100$ GeV the result of the full calculation is an order of 
magnitude larger than the fragmentation calculation, even for large $P_T$. 
However, when $\sqrt{\hat s}$ becomes extremely large; 
e.g., $\sqrt{\hat s}=800$ 
GeV, the contribution of this subset II  is dominated by the $c$ quark 
fragmentating into the $B_c$ meson when $P_T\gg M_{B_c}$. This implies that 
the higher twist terms must have very large coefficients, which violates 
that naive power counting rules which imply that the higher twist terms are 
suppressed by $\displaystyle {M_{B_c}^2\over \hat{s}}$ and are therefore 
negligible when $\sqrt{\hat s}\sim 100$ GeV $\gg M_{B_c}$. Fortunately, 
this type of contribution, which involves $b\bar{b}$ quark pair creation, 
is not important in the hadronic production of the $B_c$ and $B_c^*$ simply 
because the factor of 16 enhancement due to the larger $c$ quark electric 
charge is absent.

Having investigated the subprocess, we next calculated  $B_c$ and $B_c^*$
production at $p\bar{p}$ colliders, particularly the Fermilab Tevatron. 
From our above analysis when $P_T\gg M_{B_c}$ and, therefore,
$\sqrt{\hat{s}} \gg M_{B_c}$ the process is $\bar{b}$ quark fragmentation 
dominated. However at smaller $P_T$ the fragmentation approximation 
breaks down. In Fig. 2, we compare the $P_T$ distributions of the $B_c$ 
and the $B_c^*$ mesons coming from the full $\alpha_s^4$ calculation with 
the fragmentation approximation. In these calculations we used the CTEQ3M 
parton distribution functions\cite{16}. There is some ambiguity in the 
choice of the energy scale $\mu$ in $\alpha_s(\mu)$ and this is clearly 
dependent on  the form of the factorization. For example, the 
component of the $\bar{b}$ fragmentating into the $B_c$ can be factorized 
with a $b\bar{b}$ pair first being created at a very short distance, 
with a reasonable choice of this scale being $\mu_1=\sqrt{m_b^2+P_{Tb}^2}$, 
followed by the $\bar{b}$ fragmentating into a $B_c$ or $B_c^*$, with a 
reasonable choice of this energy scale being $\mu_2=2m_c$.
Factorization in the non-fragmentation component is more complicated 
and the choice of energy scales is not obvious. For 
simplicity, we will use a uniform choice of energy scales,  
choosing the same scale as set in  the $\bar{b}$ fragmentation
component; i.e. $\alpha_s^2(\mu_1)\times \alpha_s^2(\mu_2) $.
From the numerical calculations, Fig. 2, we see that for the $B_c$ meson 
the $P_T$ distributions for $P_T>5$ GeV are very similar for the full 
$\alpha_s^4$ calculations and the fragmentation approximation.
This critical value of $P_T$ is  much smaller than that found above in the 
study of the subprocess.  However, for the $B_c^*$ meson, the result 
predicted by the fragmentation approximation in Fig. 2 differs from that 
predicted by the full $\alpha_s^4$ calculation by 50-70\% over a much
larger range of $P_T$, more consistent with what was found in the study
of the subprocess.
Therefore, it is difficult to decide from only 
the $P_T$ distributions where the fragmentation approximation is reliable 
for the hadronic production of the $B_c$ and $B_c^*$ mesons. In fact,
we found that the $P_T$ distribution alone can be misleading.  

To clarify this issue, we introduce the distribution
\begin{equation}
C(z) \displaystyle = \int dx_1 dx_2 g(x_1,\mu) 
g(x_2,\mu) {d\hat{\sigma}(\sqrt{\hat s},\mu) \over d z},
\end{equation}
where $\displaystyle z\equiv {2(k_1+k_2)\cdot p \over {\hat{s}}}$
and $g(x_i,\mu)$ are the gluon distribution functions.
In the subprocess center of mass $z$ is simply twice the fraction
of the total energy carried by the $B_c$ or $B_c^*$ meson. 
The distribution $C(z)$ provides a sensitive means to investigate
the dynamics of the production process and the fragmentation approximation. 
Clearly, if the fragmentation approximation is valid, 
$\displaystyle {d\hat{\sigma}(\sqrt{\hat s},\mu ) \over d z}$ 
can be factorized  as 
\begin{equation}
\displaystyle {d\hat{\sigma}(\sqrt{\hat s},\mu) \over d z} =
\sum_i \hat{\sigma}_{gg\to Q_i\bar{Q}_i}\otimes  
D_{Q_i\to B_c}(z,\mu).
\end{equation}
where $D_{Q_i\to B_c}(z,\mu)$ are the usual fragmentation functions
and $\hat{\sigma}_{gg\to Q_i\bar{Q}_i}$ is the gluon fusion subprocess
cross section for production of the heavy quark pair $Q_i\bar{Q}_i$.
In this approximation, the integrals over $x_1$ and $x_2$ can be performed, the 
fragmentation function can be factored out, and $C(z)$ 
is simply proportional to a sum of the usual fragmentation functions
which is insensitive to the parton distribution functions 
and to the kinematic cuts.
However, if the distribution $C(z)$ is quite different from the 
fragmentation functions, it is an 
indication that the fragmentation approximation is not valid. Therefore, 
comparing $C(z)$ calculated in the fragmentation approximation with 
the full order $\alpha_s^4$ calculation, provides a quantitative criterion 
to judge the validity of the fragmentation approximation. We emphasize that 
$z$ is a very useful variable and is an experimentally measurable quantity, 
at least in principle.

In Fig 3 we compare the $z$ distribution $C(z)$ calculated in the 
fragmentation approximation with the full order $\alpha_s^4$
calculation for $B_c$ and $B_c^*$ meson production with a cut of $P_T>10$ GeV
(Fig. 3a) and also with cuts of $P_T>20$ GeV and $P_T>30$ GeV (Fig. 3b).
The $z$ distribution $C(z)$ is sensitive to 
the smallest $P_T$ region for a given 
$P_T$ cut because the $P_T$ distributions of the $B_c$ and $B_c^*$ mesons 
decrease very rapidly with increasing $P_T$.  From Fig. 3. some general 
features 
are evident: For the $B_c$ meson, in the fragmentation approximation,
for smaller $P_T$ cuts the $z$ distributions $C(z)$ in 
the higher $z$ region is overestimated 
while it is underestimated in the lower $z$ region;  but, after integration 
over $z$, 
the result is similar to the full $\alpha_s^4$ calculation. However, 
for the $B_c^*$ 
meson, even for the largest $P_T$ cut, 
the $z$ distribution $C(z)$ calculated in the fragmentation 
approximation is underestimated at all values of $z$ and, after integration 
over $z$, the 
result is definitely smaller than the full $\alpha_s^4$ calculation. This 
feature explains why the $P_T$ distributions shown in Fig. 2 
are similar for the $B_c$ 
meson even down to $P_T\sim M_{B_c} $
but are different for the $B_c^*$ meson. 
This also shows that it is simply fortuitous that the $P_T$ distribution 
of the $B_c$ calculated in the fragmentation approximation is similar to 
that from the full $\alpha_s^4$ calculation for $P_T$ below this value,
particularly down to $P_T\sim M_{B_c}$.
 It is also clear that when $P_T$ is increased  
the distributions become closer.
As shown in Fig. 3b, when $P_T$ is as large as 30 GeV the curves calculated
in the fragmentation approximation are quite close to the full $\alpha_s^4$
calculation. This indicates that the fragmentation 
approximation is valid in the large $P_T$ region, as expected. 
We emphasize here that the 
difference between the fragmentation approximation and full calculation is 
not universal, but is process--dependent, and thus does not satisfy 
Atarelli-Parisi evolution.

 Finally, we examined the ratio of cross sections for $B_c^*$ production
calculated in the fragmentation approximation with the results of the full 
$\alpha_s^4$ calculation. As 
discussed above, for $B_c^*$ meson production the fragmentation 
approximation always underestimates the full $\alpha_s^4$. The deviation 
from the full calculation for the $B_c^*$ meson can be used as a criterion 
to test the validity of the fragmentation approximation. The results for 
the total cross section $\sigma(P_T> P_{T \min})$ for various $P_T$ cuts  
are listed in  Table I. 
Taking agreement within $30\%$ as the criterion for the validity of the 
fragmentation approximation we also see that $P_T$ should exceed about 
30 GeV, a value considerably larger than the heavy quark masses. We also
compared the ratio of $B_c^*$ to $B_c$ production predicted by the full 
$\alpha_s^4$ calculation with the fragmentation approximation and reached 
the same conclusion.  
We note that this conclusion is also rather insensitive to the choice 
the energy scale $\mu$ and the parton distribution functions.

Both the spectroscopy and the decays of the $B$ mesons have been widely 
studied\cite{17,52,53,54}. The excited states below the threshold will 
decay to the ground state $B_c$ by emitting the photon(s) or $\pi$ mesons.  
The golden channel to detect the $B_c$ meson is  $B_c\longrightarrow J/\psi
+\pi (\rho)$. However, the branching  ratio is quite small, around 0.2. 
The exclusive semileptonic decay mode $B_c\longrightarrow J/\psi +l+\nu_l$  
has a relatively larger branch ratio, but there is ``missing energy''.
Our calculations should be helpful in searching for the $B_c$ since 
the $P_T$ distribution falls so rapidly that the small $P_T$ region is
quite important.

In summary, we  carried out a detailed comparative study of the fragmentation 
approximation and the full order $\alpha_s^4$ QCD perturbative calculation. 
We first studied the subprocess and obtained some insight into the process
by analyzing the singularities appearing in the amplitude and the $P_T$
distribution of the subprocess. This revealed that there are Feynman diagrams 
present in the full order $\alpha_s^4$ matrix element in which 
{\em two or more}
quarks or gluons can be {\em simultaneously} nearly on-mass shell and that  
these can dominate the subprocess over the fragmentation approximation in 
certain regions of the phase space. We then calculated the hadronic 
production process and investigated the $z$ distribution, $C(z)$, 
which was used to test the fragmentation approximation. From the study of 
both the subprocess and the hadronic process  we conclude that the 
fragmentation mechanism dominates when and only when $P_T\gg M_{B_c}$. 
This conclusion is independent of the choice of the energy scales and the 
parton distribution functions. It is only fortuitous that the $P_T$ 
distribution of the $B_c$ in the fragmentation approximation is similar to 
that of the full calculation for $P_T$ as low as $M_{B_c}$. 

This work was supported in part by the U.S. Department of Energy, Division 
of High Energy Physics, under Grant DE-FG02-91-ER40684 and by the  National 
Nature Science Foundation of China.

\vfill\eject

\parbox{5.8in}{ Table  I. Total cross sections 
$\sigma(P_{T\,B_c}> P_{T\,min})$  in $nb$ for hadronic 
production of the $B_c$ and the $B_c^*$ mesons predicted  by the 
$\alpha_s^4$ calculation and the fragmentation approximation assuming
various $P_T$ cuts and $|Y|<1.5$. 
The CTEQ3M parton distribution functions were used and the values
$f_{B_c}=480$ MeV, $m_c=1.5$ GeV, $m_b=4.9$
GeV, and $M_{B_c}=6.4$ GeV.}
\bigskip

\begin{center}
  \begin{tabular}{l|ccccccc} 
\hline\hline
 $P_{T\,min}$ (GeV) ~&  \makebox[1.2cm]{0} & \makebox[1.2cm]{5} 
 & \makebox[1.2cm]{10}& \makebox[1.3cm]{15}& \makebox[2.1cm]{20}&  
   \makebox[2.1cm]{25}& \makebox[2.1cm]{30}  \\ \hline 
 $\sigma_{B_c}(\alpha_s^4)$ & 1.8 & 0.57 & 0.087 & 0.018 & $4.8\times10^{-3}$
          & $1.6\times10^{-3}$  & $6.3\times10^{-4}$ \\
 $\sigma_{B_c}(\,frag.\,) $ & 1.4 & 0.47 & 0.071 & 0.014 & $4.0\times10^{-3}$
          & $1.3\times10^{-3}$ & $5.3\times10^{-4}$  \\
 $\sigma_{B_c^*}(\alpha_s^4)$ & 4.4 & 1.4 & 0.22 & 0.041 & $1.1\times10^{-2}$
          & $3.4\times10^{-3}$ & $1.3\times10^{-3}$ \\
 $\sigma_{B_c^*}(\,frag.\,) $ & 2.3 & 0.78 & 0.12 &0.025 & $6.8\times10^{-3}$
          & $2.3\times10^{-3}$ & $9.2\times10^{-4}$  \\[2mm]
 $\displaystyle
  {\sigma_{B_c^*}(\,frag.\,) \over  \sigma_{B_c^*}(\alpha_s^4) }$ 
        & 0.52 & 0.55 & 0.56 & 0.61 & 0.63 & 0.67 & 0.70 
\\ [4mm]
\hline \hline \end{tabular} 
\end{center}

\vfill\eject


\figure{Fig. 1. The $P_T$ distributions of the $B_c$ and $B_c^*$ meson for 
the subprocess with $\sqrt{\hat{s}}=$ 200 GeV. The solid and the doted 
lines correspond to the full $\alpha_s^4$ calculation and the fragmentation 
approximation, respectively.}

\figure{Fig. 2. The $P_T$ distributions of $B_c$ and $B_c^*$ meson 
production at the Tevatron energy $\sqrt{{s}}=$ 1.8 TeV. The solid  
and the doted lines correspond to the full $\alpha_s^4$ calculation 
and the fragmentation approximation, respectively. }

\figure{Fig. 3. The $z$ distributions $C(z)$ of the $B_c$ and $B_c^*$ 
at the Tevatron energy $\sqrt{s}=1.8$ TeV. The solid lines are the
full $\alpha_s^4$ calculation and the doted lines are the fragmentation
approximation (a) with the cut $P_T> 10$ GeV  and  (b) with the cuts 
$P_T> 20$ GeV and $P_T> 30$ GeV. }

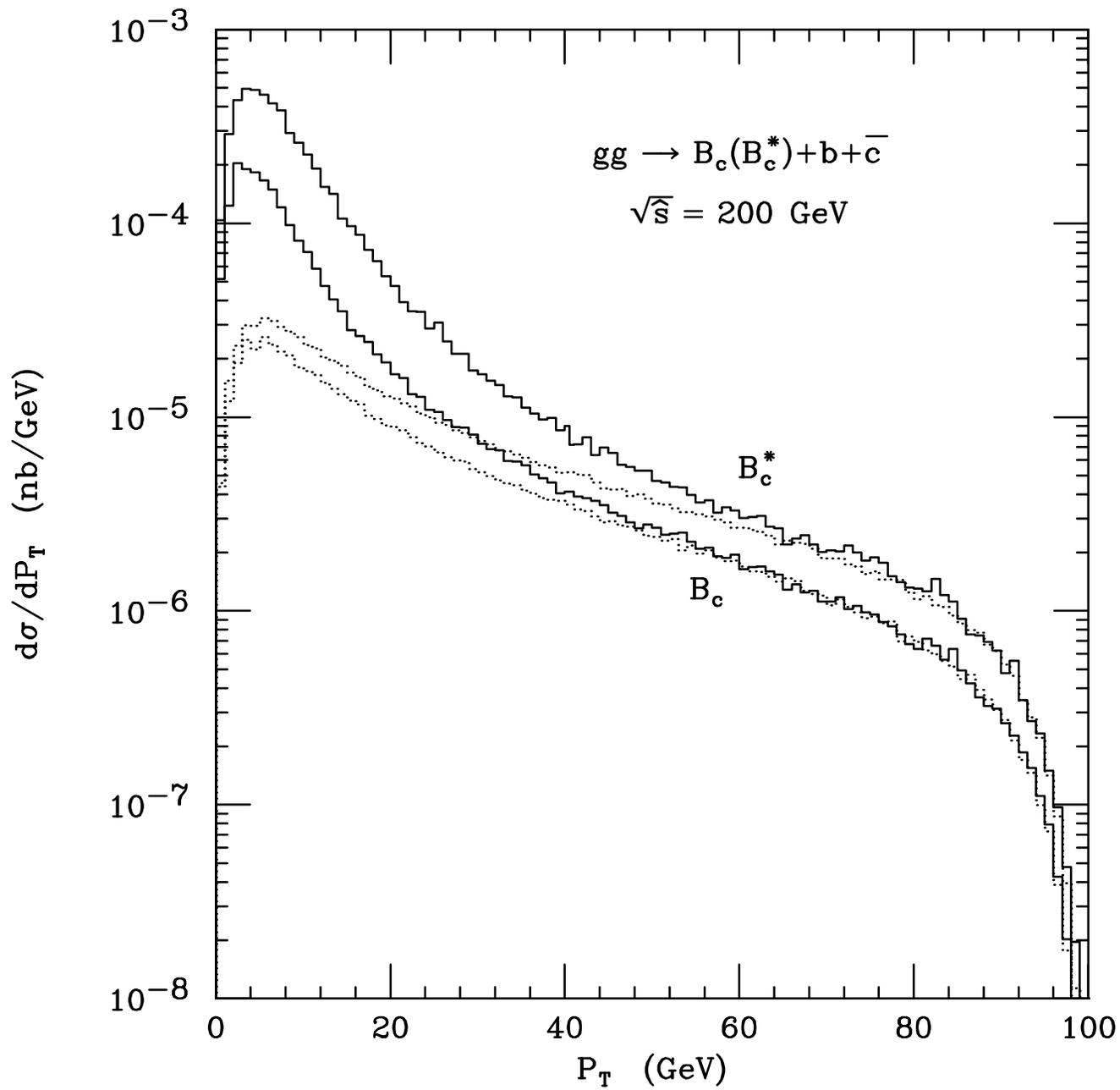
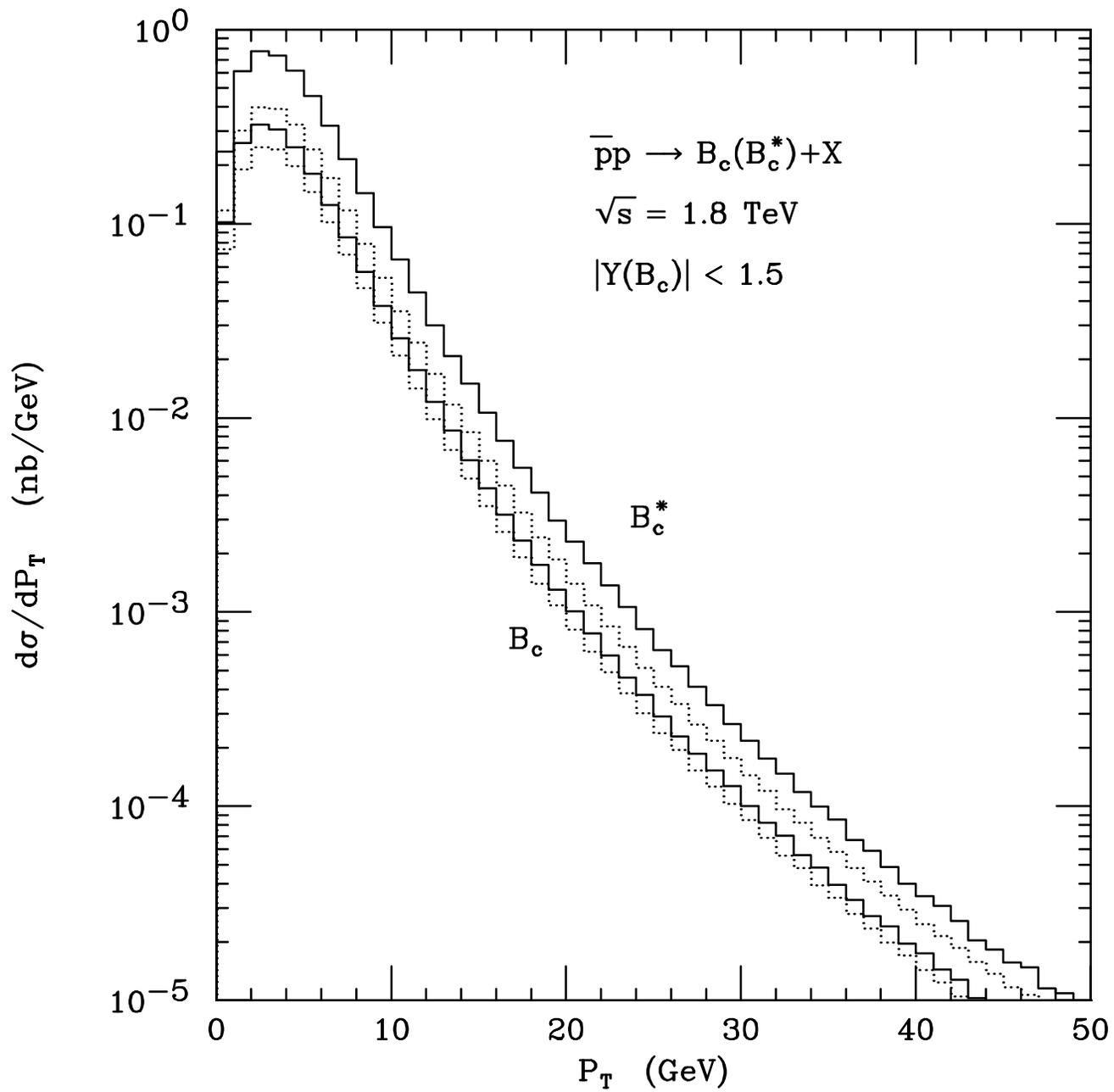
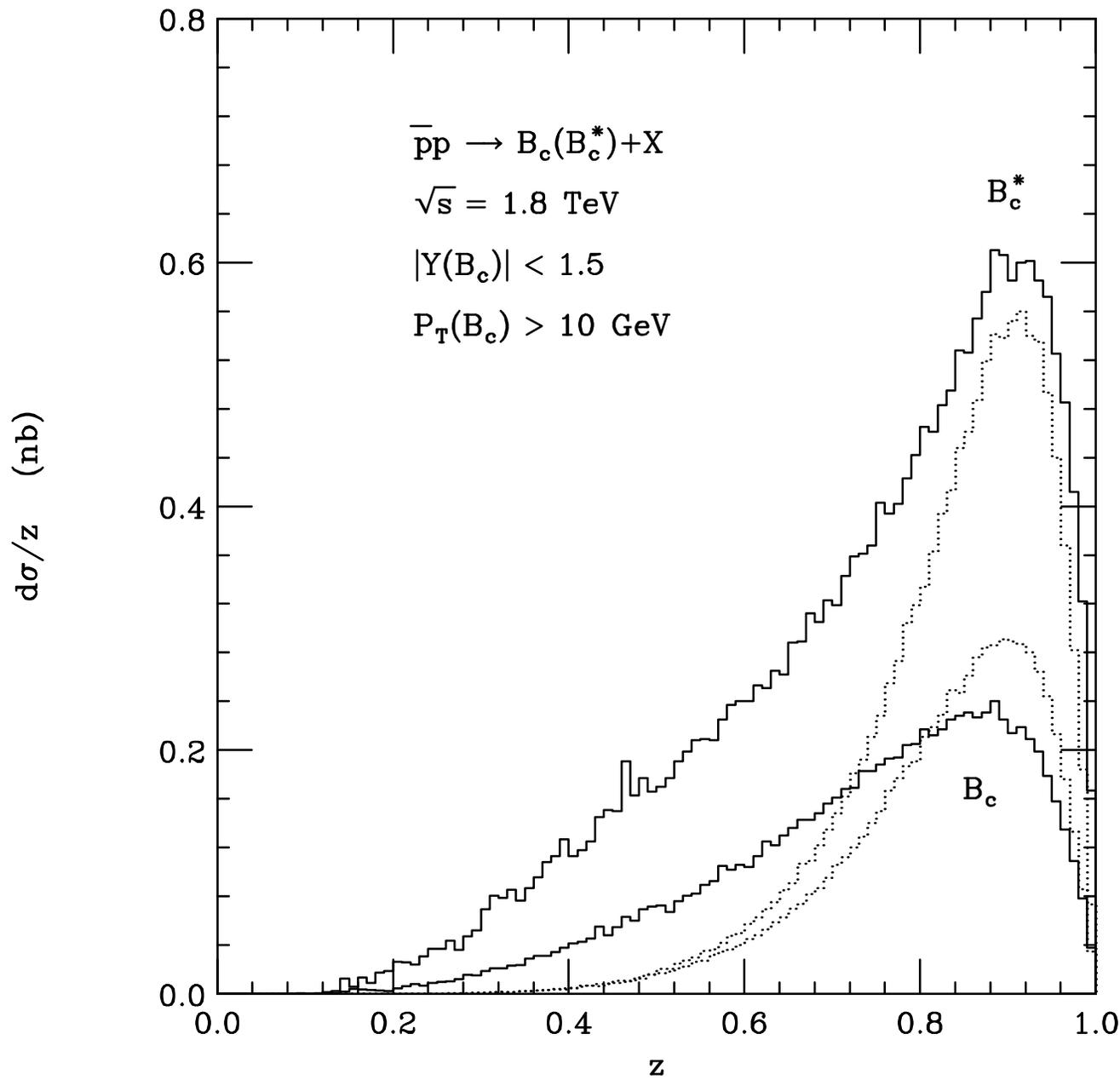
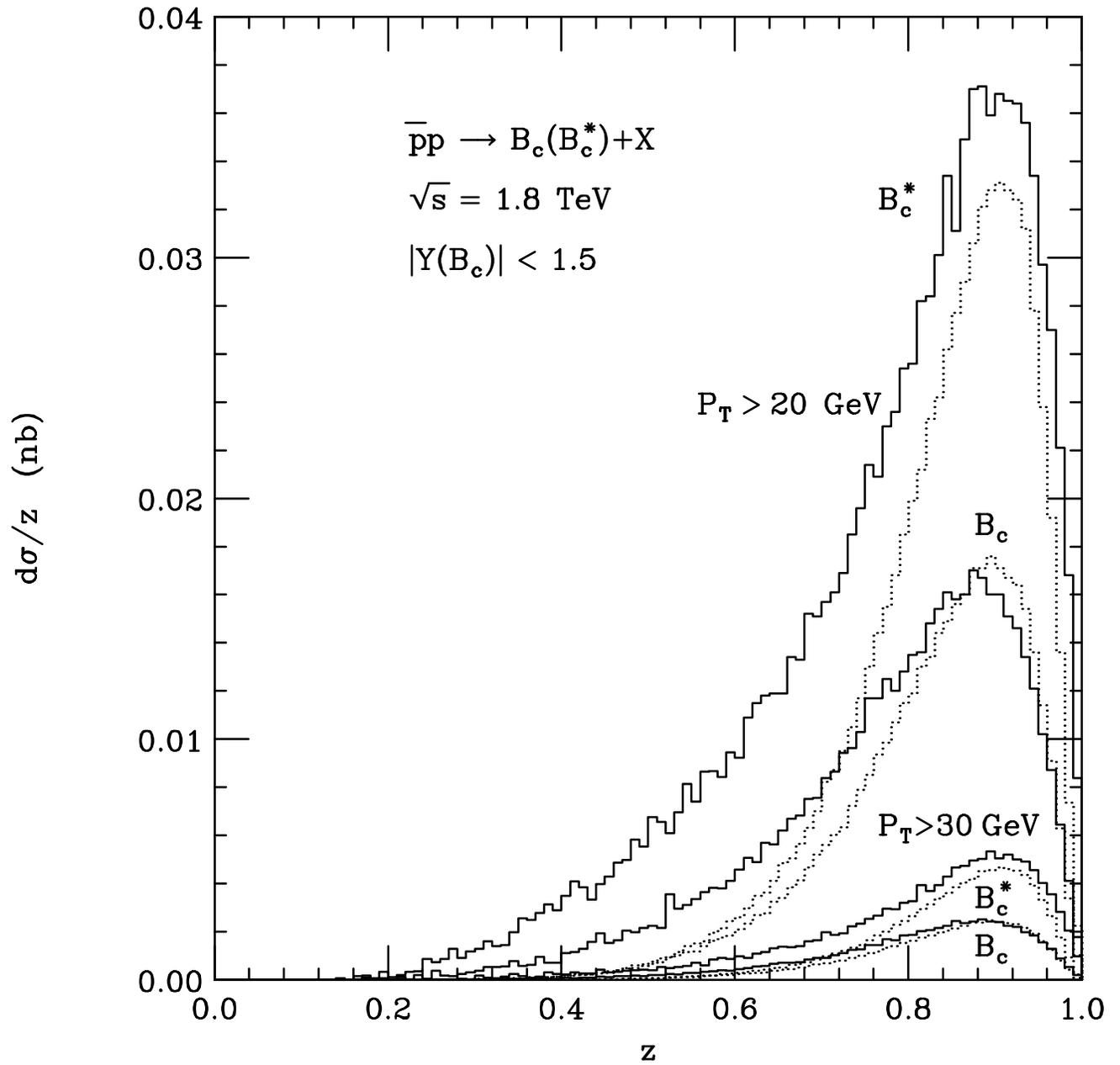
\end{document}